\documentclass[11pt]{article}
\usepackage{jheppub_modified_green}

\usepackage{tocloft}
\usepackage{epsfig}
\usepackage{amsfonts}
\usepackage{latexsym}
\usepackage{amsmath,amssymb}
\usepackage{graphicx}

\newcommand{\nc}{\newcommand}
\nc{\rnc}{\renewcommand}
\rnc{\bigstar}{\clubsuit}
\nc{\F}{\mathcal{F}}
\nc{\vq}{\vec{q}}
\nc{\bo}{\raise-0.5mm\hbox{\Large $\Box$}}
\rnc{\d}{\mathrm{d}}
\nc{\D}{\partial}
\nc{\K}{\kappa}
\nc{\M}{\mathcal{M}}
\nc{\bK}{\bar{\K}}
\nc{\bN}{\bar{N}}
\nc{\bq}{\bar{q}}
\nc{\bp}{\bar{p}}
\nc{\vbq}{\vec{\bar{q}}}
\nc{\g}{\gamma}
\nc{\lrarrow}{\leftrightarrow}
\nc{\rg}{\sqrt{g}}
\rnc{\[}{\begin{equation}}
\rnc{\]}{\end{equation}}
\nc{\bea}{\begin{eqnarray}}
\nc{\eea}{\end{eqnarray}}
\nc{\nn}{\nonumber}
\rnc{\(}{\left(}
\rnc{\)}{\right)}
\nc{\q}{\vec{q}}
\nc{\x}{\vec{x}}
\nc{\ep}{\epsilon}
\nc{\tto}{\rightarrow}
\rnc{\inf}{\infty}
\rnc{\Re}{\mathrm{Re}}
\rnc{\Im}{\mathrm{Im}}
\nc{\z}{\zeta}
\nc{\I}{\mathcal{I}}
\nc{\mA}{\mathcal{A}}
\nc{\A}{\mA}
\nc{\mB}{\mathcal{B}}
\nc{\mC}{\mathcal{C}}
\nc{\mD}{\mathcal{D}}
\nc{\mE}{\mathcal{E}}
\nc{\mF}{\mathcal{F}}
\rnc{\H}{\mathcal{H}}
\rnc{\L}{\mathcal{L}}
\nc{\<}{\langle}
\rnc{\>}{\rangle}
\nc{\fnl}{f_{NL}}
\nc{\fnleq}{f_{NL}^{equil.}}
\nc{\fnlloc}{f_{NL}^{local}}
\nc{\vphi}{\varphi}
\nc{\Lie}{\pounds}
\nc{\half}{\frac{1}{2}}
\nc{\bOmega}{\bar{\Omega}}
\nc{\bLambda}{\bar{\Lambda}}
\nc{\dN}{\delta N}
\nc{\gYM}{g_{\mathrm{YM}}}
\nc{\geff}{g_{\mathrm{eff}}}
\nc{\bg}{\hat{\gamma}}
\nc{\tO}{\tilde{\O}}
\nc{\Oi}{\Omega_{[2]}}
\nc{\Oii}{\Omega_{[3]}}
\nc{\Ei}{E_{[2]}}
\nc{\Eii}{E_{[3]}}
\nc{\bOi}{\bar{\Omega}_{[2]}}
\nc{\bOii}{\bar{\Omega}_{[3]}}
\nc{\bEi}{\bar{E}_{[2]}}
\nc{\bEii}{\bar{E}_{[3]}}
\rnc{\a}{\bar{a}}
\rnc{\b}{\bar{b}}
\rnc{\c}{\bar{c}}
\rnc{\O}{\mathcal{O}}
\nc{\blambda}{\bar{\lambda}}
\nc{\oa}{\stackrel{\leftrightarrow}}
\newcommand{\lla}{\langle \! \langle}
\newcommand{\rra}{\rangle \! \rangle}

\newcommand{\p}{\partial}
\nc{\wT}{\widetilde{T}}
\rnc{\O}{\mathcal{O}}
\rnc{\L}{\mathcal{L}}
\nc{\T}{\mathcal{T}}
\nc{\Y}{\mathcal{Y}}
\nc{\X}{\mathcal{X}}
 
 \title{On the power spectrum of inflationary cosmologies dual to a deformed CFT} 
\author{Paul McFadden}
\affiliation{Perimeter Institute for Theoretical Physics,\\ Waterloo, Ontario, Canada N2L 2Y5.}
\emailAdd{pmcfadden@perimeterinstitute.ca}
\abstract{
We analyse slow-roll inflationary cosmologies that are holographically dual to a 
three-dimensional conformal field theory deformed by a nearly marginal scalar operator.
We show the cosmological power spectrum 
is inversely proportional to the spectral density associated with the 2-point function of the trace of the stress tensor in the deformed CFT.  
Computing this quantity using second-order conformal perturbation theory, 
we obtain a holographic power spectrum in exact agreement  
with the expected inflationary power spectrum to second order in slow roll.
}
\arxivnumber{}

\begin{document}

\maketitle

\section{Introduction}

In this paper we pursue 
precision holographic cosmology. 
Starting from a three-dimensional quantum field theory -- specifically, the deformation of a conformal field theory by a nearly marginal scalar operator -- we compute {\it holographically} the power spectrum of the dual inflationary cosmology to second order in slow roll.
Our strategy is extremely simple.
First, we show that the cosmological power spectrum is inversely proportional to a certain spectral density in the dual QFT, namely that associated with the 2-point function for the trace of the stress tensor.
We then compute this quantity directly
in the dual QFT using conformal perturbation theory, essentially an expansion in the small parameter 
controlling the nearly marginal dimension of the scalar operator.
From a cosmological perspective, this scalar operator is dual to the inflaton, and our procedure is analogous to an expansion in the spectral tilt.  Crucially, this expansion is valid even when the dual QFT is strongly coupled, as is the case when the bulk inflationary cosmology is governed by ordinary Einstein gravity.
In this sense, we are effectively able to compute on both sides of the holographic correspondence simultaneously, setting up a striking test of holographic cosmology.

Our scenario is then as follows.  
Driven by the nearly marginal scalar operator, the dual QFT undergoes an RG flow from the original UV CFT to a nearby IR fixed point.  Since the two fixed points lie close together, the coupling of the scalar operator never becomes large along the flow, and in fact is bounded by the same small parameter controlling the dimension of the operator.  Cosmologically, each fixed point of the RG flow corresponds to a de Sitter asymptotic region, since the three-dimensional conformal group (in Euclidean signature) coincides with the isometry group of four-dimensional de Sitter spacetime.  We thus have an inflationary cosmology undergoing a slow-roll evolution from an early-time de Sitter phase, corresponding to the IR fixed point, to a different late-time de Sitter phase governed by the UV fixed point.  The inflaton rolls from a local maximum of the potential down to a  nearby local minimum, with the total distance traversed remaining small.  (We will not consider 
 how to exit inflation here, though this may be accomplished by allowing additional operators to enter the RG flow  changing its fate in the UV.)  

Since the slow-roll parameter $\ep$ vanishes at each extremum of the potential, and the separation of these extrema is moreover small, we find ourselves in a situation where the slow-roll parameters 
satisfy $\ep\ll \eta$.  
Our holographic calculation in the dual QFT thus successfully recovers the 
slow-roll inflationary power spectrum up to and including the second-order
$\eta^2$ and $\delta_2$ corrections, including all the correct numerical prefactors.  
(To recover the corrections proportional to $\ep$ would require working to 
higher order still 
in conformal perturbation theory.)
As well as offering a fresh perspective on the inflationary 
power spectrum, these results provide  
a remarkable confirmation of the consistency of the holographic framework.

Our analysis in this paper refines and extends 
that of \cite{Bzowski:2012ih}, which tackles both the power spectra and non-Gaussianities for the same scenario, but restricted to leading order in conformal perturbation theory. 
 The second-order analysis we present here resolves a number of important 
 technical issues that could be avoided at leading order, most notably the definition of the renormalised scalar operator away from criticality.  Here, we define
the scalar operator so that the corresponding running coupling exactly coincides with 
the bulk inflaton 
at horizon crossing.
Our present approach is also significantly 
streamlined. 
By using RG-improved perturbation theory, 
we are able to dispense with the explicit
series resummation of correlators with integrated insertions performed in  \cite{Bzowski:2012ih}.  We also now derive the holographic power spectrum directly in terms of the slow-roll parameters at horizon crossing, without requiring their evaluation as specific functions of momentum.

Related recent discussions of holographic cosmologies dual to a deformed CFT may be found in \cite{Schalm:2012pi, Mata:2012bx, Garriga:2013rpa},
while relevant earlier literature inspired by the 
dS/CFT correspondence \cite{Strominger:2001pn,Witten:2001kn,Maldacena:2002vr}
includes 
\cite{
Strominger:2001gp, Larsen:2002et, Halyo:2002zg, Larsen:2003pf, vanderSchaar:2003sz, Larsen:2004kf, Seery:2006tq}.
For a pure CFT calculation for the inflationary correlator of three gravitons, see \cite{ Maldacena:2011nz, Bzowski:2011ab}, 
and for other recent approaches to holographic cosmology, see 
\cite{
Antoniadis:2011ib,
Harlow:2011ke, Creminelli:2011mw,
Dong:2011uf, Anninos:2011ui, 
Hertog:2011ky, 
Kehagias:2012pd,
Castro:2012gc,Smolkin:2012er,
Anninos:2013rza,
Hartle:2013vta,  
Banks:2013qra}.

The outline of 
this paper is as follows.  In Section \ref{sec:spec_fns}, we introduce the relevant spectral representation for the stress tensor 2-point function in the dual QFT and discuss its relation to the power spectra of the corresponding cosmology.  In Section \ref{sec:CPT} we turn to conformal perturbation theory.  We discuss the deformation of a three-dimensional CFT by a nearly marginal scalar operator, computing the $\beta$-function of the deformed theory and the RG-improved 2-point function for the trace of the stress tensor.  We then extract the spectral density with the help of a dispersion relation.  In Section \ref{AdSNorms} we consider the nature of the UV CFT, using simple AdS/CFT calculations on a non-dynamical AdS background to evaluate the required input data for our conformal perturbation theory.  
Section \ref{sec:calibrating} then discusses how to choose a renormalisation scheme so that the bulk inflaton at horizon crossing maps precisely to the running coupling in the dual QFT.  This choice of scheme simplifies the translation of QFT variables into cosmological parameters, with the $\beta$-function for the running coupling and its derivatives mapping to the bulk slow-roll parameters at horizon crossing.
Our result for the holographic power spectrum follows in Section \ref{sec:results}, where we elaborate on the nature of the bulk cosmology.  We conclude in Section \ref{sec:discussion} with a discussion of open directions.  
Two appendices provide additional 
technical information: Appendix \ref{app:B} solves for the running coupling in the dual QFT and verifies the consistency of our results with those of \cite{Bzowski:2012ih},
while Appendix \ref{WittenDias} presents 
details of 
the AdS/CFT calculation of correlators 
summarised in Section \ref{AdSNorms}.

\section{Holographic power spectra from spectral functions}
\label{sec:spec_fns}

We begin in this first section 
with a general discussion of the holographic power spectrum.
Our approach to holographic cosmology is based on standard AdS/CFT  
in combination with the domain-wall/cosmology correspondence.  
In \cite{McFadden:2009fg, McFadden:2010na}, holographic formulae were derived for the primordial scalar and tensor power spectra in terms of the 2-point function for the stress tensor in the dual QFT.  
(For higher-point correlation functions, see \cite{McFadden:2010vh,McFadden:2011kk}.)
This dual QFT is three-dimensional and it is convenient  
to take it to live in Euclidean signature.
  
On general grounds, in momentum space the stress tensor 2-point function takes the form
\[
\lla T_{ij}(q)T_{kl}(-q)\rra = A(q^2)\Pi_{ijkl}+B(q^2)\pi_{ij}\pi_{kl},
\]
where the transverse traceless and transverse projectors 
\[
\Pi_{ijkl} = \pi_{i(j}\pi_{k)l}-(1/2)\pi_{ij}\pi_{kl}, \qquad \pi_{ij}=\delta_{ij}-q_iq_j/q^2. 
\]
The function $A(q^2)$ thus encodes the transverse traceless piece of the correlator, while $B(q^2)$ encodes its trace, namely $\lla T(q)T(-q)\rra = 4B(q^2)$.
Our double bracket notation here simply indicates the removal of the momentum-conserving delta function, i.e.,
\[
\<T_{ij}(q_1)T_{kl}(q_2)\> = (2\pi)^3\delta(q_1+q_2)\lla T_{ij}(q)T_{kl}(-q)\rra .
\]

Through standard AdS/CFT, 
the stress tensor 2-point function for a three-dimensional QFT flowing to a fixed point in the UV 
can be related to the asymptotic behaviour of metric perturbations in a four-dimensional asymptotically AdS domain-wall spacetime.  
The domain-wall/cosmology correspondence then allows these domain-wall fluctuations to be mapped to cosmological perturbations on a corresponding cosmological spacetime. 
In this manner the stress tensor 2-point function in the dual QFT can be 
related to the late-time scalar and tensor power spectra in the corresponding cosmology, $\Delta_S^2(q)$ and $\Delta_T^2(q)$, by the following holographic formulae:
\[\label{hol_form_1}
\Delta_S^2(q) = \frac{1}{16\pi^2}\frac{q^3}{\Im\,B(q^2)}, \qquad \Delta_T^2(q) = \frac{2}{\pi^2}\frac{q^3}{\Im\,A(q^2)}.
\]
The imaginary parts in these expressions are taken after applying the analytic continuation $q\tto -iq$, where  $q = +\sqrt{q^2}$ is the momentum magnitude.  (The 3-momentum in the dual QFT corresponds to the comoving 3-momentum on spatial slices in the cosmology.)

In fact, the formulae quoted in \cite{McFadden:2009fg, McFadden:2010na} contain an additional minus sign associated with a continuation $N^2\tto -N^2$ of the rank of the gauge group of the dual QFT.  
Since the only appearance of $N^2$ here is as an overall factor multiplying $A(q^2)$ and $B(q^2)$, however,
we have simply included the corresponding sign directly into the holographic formulae above.

In the following, we wish to suggest an alternative physical interpretation for these formulae in terms of the spectral representation for the stress tensor 2-point function.
In three Euclidean dimensions, adopting the conventions of \cite{Cappelli:1990yc}, this reads
\[
\<T_{ij}(x)T_{kl}(0)\> = \frac{\pi}{16}\int_0^\infty\d\rho\,\Big[\frac{1}{2}c^{(0)}(\rho)S_{ij}S_{kl}+c^{(2)}(\rho)(2S_{i(j}S_{k)l}-S_{ij}S_{kl})\Big]\int\frac{\d^3 q}{(2\pi)^3}\frac{e^{iq\cdot x}}{q^2+\rho^2},
\]
where $S_{ij}=\p_i\p_j-\delta_{ij}\p^2$.  Physically, the dimensionless spectral functions $c^{(0)}(\rho)$ and $c^{(2)}(\rho)$ encode the contribution to the 2-point function from  spin-0 and spin-2 intermediate states of mass $\rho$.  Originally, interest in these spectral functions was motivated 
by attempts to extend Zamolodchikov's c-theorem to higher dimensions, see, e.g., \cite{Shore:1990wq,Cappelli:1990yc,Shore:1990ng,Cappelli:1991ke}.

\begin{figure}[t]
\center
\includegraphics[height=4.7cm]{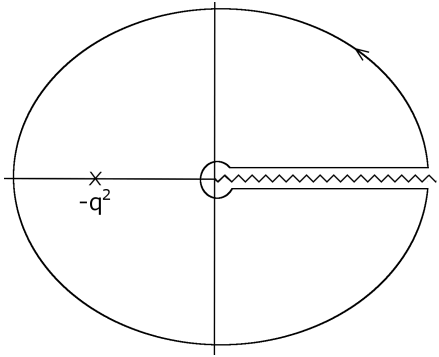} 
\caption{\label{Branch} To derive the dispersion relation \eqref{disp2} we integrate $B(-\rho^2)/(q^2+\rho^2)^3$ over the contour in the complex $\rho^2$ plane shown above.  The contribution from the circular arc at infinity vanishes, since in the case of interest, $B(\rho^2) \tto \rho^{3-2\lambda}$ as $\rho\tto \infty$ with $0<\lambda\ll 1$.  The discontinuity across the branch cut
is then related to the residue at $\rho^2=-q^2$.}
\end{figure}

In momentum space, the spectral representation takes the form
\[\label{mom_repAB}
A(q^2) = \frac{\pi}{8}\int_0^\infty\d\rho\,c^{(2)}(\rho)\frac{q^4}{q^2+\rho^2}, \qquad B(q^2) = \frac{\pi}{32}\int_0^\infty\d\rho\,c^{(0)}(\rho)\frac{q^4}{q^2+\rho^2}.
\]
These formulae allow $A(q^2)$ and $B(q^2)$ to be analytically continued to the complex $q^2$ plane with a branch cut along the negative real axis. 
The spectral functions $c^{(0)}(\rho)$ and $c^{(2)}(\rho)$ may then be extracted by applying standard dispersion relations \cite{Cappelli:1990yc}.  For example, integrating along the contour given in Figure \ref{Branch}, we find
\[\label{disp2}
B''(q^2) = 2\int_0^\infty \frac{\d\rho^2}{\pi} \frac{\Im\, B(-\rho^2-i\ep)}{(q^2+\rho^2)^3}, \qquad \ep\tto 0_+.
\]
Differentiating \eqref{mom_repAB} directly, on the other hand, yields
\[
B''(q^2) = \frac{\pi}{32}\int_0^\infty \d\rho^2\, c(\rho)\frac{\rho^3}{(q^2+\rho^2)^3}.
\]
In this fashion, we identify 
\[\label{c0disp}
c^{(0)}(\rho)= \frac{64}{\pi^2}\frac{1}{\rho^3}\Im\,B(-\rho^2-i\ep), \qquad
c^{(2)}(\rho)= \frac{16}{\pi^2}\frac{1}{\rho^3}\Im\,A(-\rho^2-i\ep).
\]
Given the branch cut lies along the negative real $q^2$ axis, the imaginary parts in these formulae are precisely the same as those in the holographic formulae \eqref{hol_form_1}, and hence 
\[\label{hol_form_2}
\boxed{\Delta_S^2(q) = \frac{4}{\pi^4} \frac{1}{c^{(0)}(q)}, \qquad \Delta_T^2(q) = \frac{32}{\pi^4}\frac{1}{c^{(2)}(q)}.}
\]
Expressed in this form, the holographic formulae relate physical quantities in both theories -- 
the cosmological power spectra and the spectral functions in the dual QFT -- without any analytic continuations.
Since we have incorporated the $N^2\tto-N^2$ continuation of \cite{McFadden:2009fg, McFadden:2010na} into the holographic formulae \eqref{hol_form_1} (rather than than applying it to the dual QFT), the dual QFT here is unitary meaning the spectral functions above are positive \cite{Cappelli:1990yc}.

\section{Conformal perturbation theory}
\label{sec:CPT}

Our goal in this section 
is now to compute the spectral function $c^{(0)}(q)$ for a CFT perturbed by a nearly marginal scalar operator, using RG-improved conformal perturbation theory.  
Later we will substitute our result into the holographic formula \eqref{hol_form_2} to recover the slow-roll inflationary power spectrum for scalar perturbations.
In this paper we will focus solely on the scalar power spectrum, since the tensor power spectrum 
is simply a constant at the order to which we will work \cite{Bzowski:2012ih}.

We have split this section into three parts.  First, in Section \ref{sec:defCFT}, we discuss the calculation of correlators in the deformed CFT, our choice of renormalisation scheme,
and the computation of the $\beta$-function.   
Then, in Section \ref{sec:RGimp}, we use the renormalisation group to improve our perturbative calculation of 2-point functions in the deformed CFT, to recover the correct scaling behaviour in the IR.  Finally, in Section \ref{sec:specfn}, we extract the spectral function $c^{(0)}(q)$ using the dispersion relation \eqref{c0disp}.
Our treatment of conformal perturbation theory is modelled on that in  \cite{Cappelli:1989yu}, see also
\cite{Zamolodchikov:1987ti, Ludwig:1987gs}. 

\subsection{Deforming a CFT} 
\label{sec:defCFT}

Consider a three-dimensional Euclidean CFT deformed by a slightly relevant scalar operator,
\[
S = S_{CFT}+\phi_0 \int\d^3 x  \O_0(x),
\]
where $\O_0$ has dimension $\Delta=3-\lambda$ with $0<\lambda\ll 1$.  
Correlators in the perturbed theory are given by the perturbative expansion
\begin{align}
\<\O_0(x_1)\O_0(x_2)\> &= \<\O_0(x_1)\O_0(x_2) \exp(-\phi_0\int\d^3z\,\O_0(z))\>_0 \nn\\
&= \<\O_0(x_1)\O_0(x_2)\>_0 -\phi_0 \int\d^3 z_1 \<\O_0(x_1)\O_0(x_2)\O_0(z_1)\>_0 \nn\\
&\qquad +\frac{1}{2}\phi_0^2\int\d^3 z_1\int \d^3 z_2\<\O_0(x_1)\O_0(x_2)\O_0(z_1)\O_0(z_2)\>_0 + O(\phi_0^3),
\end{align}
where the zero subscript indicates the correlator evaluated in the original undeformed UV CFT.
Transforming to momentum space, the integrated insertions become insertions at zero momentum, and we obtain
\begin{align}\label{OO_raw}
\lla \O_0(q)\O_0(-q)\rra &=   A_0 q^{3-2\lambda}- A_1 \phi_0 q^{3-3\lambda}+\frac{1}{2}A_2 \phi_0^2 q^{3-4\lambda} +O(\phi_0^3),
\end{align}
where from the dilatation Ward identity\footnote{In general this identity may be anomalous, but this is not the case here; see Section \ref{AdSNorms} and Appendix \ref{WittenDias}.} we have 
\begin{align}\label{A0def}
\lla \O_0(q)\O_0(-q)\rra_0 &= A_0 q^{3-2\lambda}, \\
\label{A1def}
\lla \O_0(q)\O_0(-q)\O_0(0)\rra_0 &= A_1 q^{3-3\lambda}, \\
\label{A2def}
\lla \O_0(q)\O_0(-q)\O_0(0)\O_0(0)\rra_0 &= A_2 q^{3-4\lambda}.
\end{align}
Our double bracket notation once again indicates the removal of the momentum-conserving delta function, e.g.,
\[
\< \O_0(q_1)\O_0(q_2)\O_0(0)\> = (2\pi)^3\delta(q_1+q_2)\lla \O_0(q)\O_0(-q)\O_0(0)\rra.
\]
The $A_n$ are numerical coefficients characterising the UV CFT: in particular, $A_0$ encodes the normalisation of $\O_0$, while $A_1$ encodes the OPE coefficient of $\O_0$ with itself. 
We will return to evaluate these coefficients in Section \ref{AdSNorms}, 
however we leave them unspecified for now. 

In the limit $\lambda\tto 0$ where $\O_0$ becomes marginal, we typically encounter singularities. 
For example, if we convert the standard position-space expression for a CFT 3-point function to momentum space, then send $\lambda\tto 0$ keeping the OPE coefficient fixed, we find $A_1 \sim \lambda^{-1}$.
To cure these divergences we introduce the dimensionless renormalised coupling $g=g(\phi_0)$ and the renormalised operator
\[\label{renOdef}
\O = \frac{\O_0}{\sqrt{Z(g)}}
\]
defined so that correlators of $\O$ are finite in the limit $\lambda\tto 0$ with $g$ fixed.  Note here that the renormalisation factor $Z(g)$ introduces a $g$ dependence into $\O$.

To define our renormalisation scheme, we  set
\[\label{scheme}
\lla \O(\mu)\O(-\mu)\rra = A_0 \mu^3
\]
at some momentum scale $\mu$, where the power of $\mu$ on the right-hand side is fixed by the engineering dimension of the correlator.  (Since $g$ is dimensionless, $\O$ has engineering dimension three.)
As usual, there is a degree of arbitrariness in any given choice of scheme: we will return to exploit this freedom later in Section \ref{sec:calibrating}.  
Solving for the renormalisation factor, we find
\[
\sqrt{Z} = \mu^{-\lambda}\Big[1-\frac{A_1}{2A_0}\phi_0\mu^{-\lambda}+\frac{1}{4}\Big(\frac{A_2}{A_0}-\frac{A_1^2}{2A_0^2}\Big)\phi_0^2\mu^{-2\lambda}+O(\phi_0^3)\Big].
\]

\begin{figure}[t]
\center
\includegraphics[height=4.0cm]{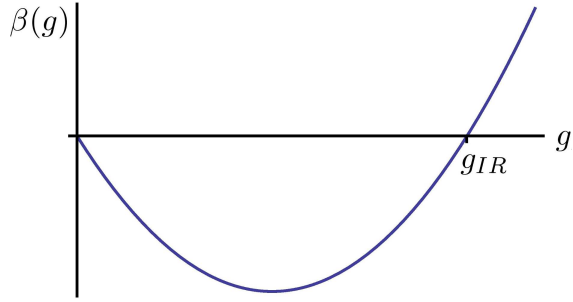} 
\caption{\label{betafig} For $b_2$ is positive and of order unity, the $\beta$-function has a perturbative IR fixed point located at $g_{IR}\sim\lambda$. }
\end{figure}

The $\beta$-function for the renormalised coupling, $\beta(g)$, is defined by 
\[
T = \beta(g) \O, \qquad \mu\frac{\d g(\mu)}{\d \mu } = \beta (g),
\]
where $T=T^i_i$ is the trace of the stress tensor.  
In the UV CFT, on the other hand, the trace of the stress tensor is 
\[
T_0 = -\lambda\phi_0 \O_0.
\]
Being a conserved current, the stress tensor does not renormalise, so $T_0=T$ and hence
\[\label{beta1}
\beta(g) = -\lambda\phi_0 \sqrt{Z} = 
-\lambda\phi_0\mu^{-\lambda}\Big[1-\frac{A_1}{2A_0}\phi_0\mu^{-\lambda}+\frac{1}{4}\Big(\frac{A_2}{A_0}-\frac{A_1^2}{2A_0^2}\Big)\phi_0^2\mu^{-2\lambda}+O(\phi_0^3)\Big].
\]
Integrating the $\beta$-function, requiring that $g(\mu)\tto \phi_0 \mu^{-\lambda}$ as $\mu\tto \infty$, then fixes
\[
g(\mu) = \phi_0\mu^{-\lambda}-\frac{A_1}{4A_0}\phi_0^2\mu^{-2\lambda}+\frac{1}{12}\Big(\frac{A_2}{A_0}-\frac{A_1^2}{2A_0^2}\Big)\phi_0^3\mu^{-3\lambda}+O(\phi_0^4).
\]
Inverting and substituting back into \eqref{beta1}, we obtain 
\begin{align}
\beta(g) &= -\lambda g + b_2 g^2+ b_3g^3 +O(g^4), 
\end{align}
where
\[ \label{b2b3_def}
b_2 = \frac{\lambda}{4}\frac{A_1}{A_0}, \qquad b_3 = \frac{\lambda}{6}\Big(\frac{5 A_1^2}{4 A_0^2}-\frac{A_2}{A_0}\Big).
\]
We will be interested in the case where $b_2$ and $b_3$ are of order unity, which requires $A_1\sim \lambda^{-1} A_0$ and $A_2\sim \lambda^{-2} A_0$ with a cancellation of the leading term $\sim \lambda^{-1}$ in $b_3$,
as we will find in Section \ref{AdSNorms}.
If $b_2$ is positive, moreover, we obtain an RG flow as illustrated in Figure \ref{betafig} to a new IR fixed point located at
\[
g_{IR} = \frac{b_2}{2b_3}\Big(-1+\sqrt{1+\frac{4\lambda b_3}{b_2^2}}\,\Big) = 
\frac{1}{b_2}\lambda-\frac{b_3}{b_2^2}\lambda^2+O(\lambda^3).
\]
With $b_2$ and $b_3$ of order unity, $g_{IR} \sim \lambda$ and the IR fixed point lies close to the UV fixed point at the origin.
The scaling dimension of $\O$ at the IR fixed point is
\[
\Delta_{IR} = 3+\beta'(g_{IR}) = 3+\lambda+\frac{b_3}{b_2^2}\lambda^2 +O(\lambda^3). 
\]

\subsection{RG-improved correlators}
\label{sec:RGimp}

Re-expressing \eqref{OO_raw} in terms of the renormalised coupling, we obtain
\begin{align}
\label{OO_raw2}
\lla \O(q)\O(-q)\rra &= A_0 q^{3-2\lambda}\mu^{2\lambda}\Big[1+\frac{4b_2}{\lambda}(1- (q/\mu)^{-\lambda})g \nn\\ &\qquad +\Big(\big(\frac{10 b_2^2}{\lambda^2}-\frac{3b_3}{\lambda}\big) (1- (q/\mu)^{-\lambda})+\frac{6 b_3}{\lambda}\Big)(1- (q/\mu)^{-\lambda}) g^2+O(g^3)  \Big].
\end{align}
This expression, however, manifestly fails to capture the expected  
$\sim q^{2\Delta_{IR}-3}$ scaling behaviour about the IR fixed point.  
To rectify this defect we turn to RG-improved perturbation theory,
implemented by solving
the Callan-Symanzik equation.  
In effect, this serves to automatically resum an 
infinite number of terms 
in the perturbative expansion.\footnote{By contrast, in Section 2.2 of \cite{Bzowski:2012ih}, this resummation was performed manually after recursively 
constructing the coefficient of each term in the series in powers of $g$.  
The differential equation that was used to construct these coefficients is equivalent to that obtained by inserting a series solution in powers of $g$ into the Callan-Symanzik equation here.
}


Since 
$
\lla \O(q)\O(-q)\rra = Z^{-1} \lla \O_0(q)\O_0(-q)\rra
$ 
and $\lla \O_0(q)\O_0(-q)\rra$ is independent of the RG scale $\mu$, we have the Callan-Symanzik equation
\[
0 = \Big(\mu\frac{\p}{\p \mu}+\beta(g)\frac{\p}{\p g} +2\gamma\Big)\lla \O(q)\O(-q)\rra,
\] 
where the anomalous dimension
\[
\gamma = \frac{1}{2}\frac{\d \ln Z}{\d \ln g} = -\lambda+2b_2 g+3b_3 g^2+O(g^3) = \frac{\d \beta}{\d g}.
\]
(Another way to arrive at this relation is to note that the stress tensor 2-point function 
$
\lla T(q)T(-q)\rra=\beta^2(g)\lla \O(q)\O(-q)\rra
$ 
has zero anomalous dimension.)

On dimensional grounds, we then have
\[
0 = \Big({-}q\frac{\p}{\p q}+\beta\frac{\p}{\p g}+3+2\gamma\Big)\lla\O(q)\O(-q)\rra.
\]
The solution takes the form
\[
\lla \O(q)\O(-q)\rra = q^3\beta^{-2}(g) \F(\bar{g}(q)),
\]
where the running coupling $\bar{g}(q)$ is defined by
\[\label{g_run}
\frac{\d\bar{g}(q)}{\d\ln (q/\mu)} = \beta(\bar{g}(q)), \qquad \bar{g}(\mu)=g.
\]
The function $\F$ then follows from our renormalisation scheme \eqref{scheme},
\[
\lla \O(\mu)\O(-\mu)\rra = A_0 \mu^3 = \mu^3 \beta^{-2}(g) \F(g) \quad \Rightarrow \quad \F(g) = A_0 \beta^2(g),
\]
hence  
\[\label{scalar2ptresult}
\lla \O(q)\O(-q)\rra = A_0 q^3 \beta^{-2}(g)\beta^2(\bar{g}(q)).
\]
We can check that the UV behaviour of this result is consistent with \eqref{OO_raw2} by solving for the running coupling about the UV fixed point as a series in $q^{-\lambda}$.
Likewise, solving for the running coupling about the IR fixed point as a series in $q^\lambda$, we now also obtain the correct scaling behaviour in the IR.

\subsection{Spectral function}
\label{sec:specfn}

We turn now to the extraction of the spectral function $c^{(0)}(q)$.
Multiplying \eqref{scalar2ptresult} by $\beta^2(g)$, the RG-improved result for the trace of the stress tensor 2-point function is 
\[
\lla T(q)T(-q)\rra 
 = A_0 q^3 \beta^2(\bar{g}(q)).
\]
The spectral function $c^{(0)}(q)$ then follows from the dispersion relation \eqref{c0disp}, 
\[
c^{(0)}(q) = \frac{16}{\pi^2}\frac{1}{q^3}\,\Im\lla T(\rho)T(-\rho)\rra\Big|_{\rho^2=-q^2-i\ep}.
\]
If an explicit solution of \eqref{g_run} for the running coupling $\bar{g}(q)$ is available (see Appendix \ref{app:B}) then this formula can be evaluated directly. 
For our present purposes, however, it is sufficient to 
write $\bar{g}(q) = G(\ln (q/\mu))$ and Taylor expand:
\begin{align}
\label{grhoeq}
\bar{g}(\rho) &= G(\ln (q/\mu) -i\pi/2) = G(\ln (q/\mu)) - \frac{i\pi}{2}G'(\ln (q/\mu))-\frac{\pi^2}{8}G''(\ln (q/\mu)) + \ldots \nn\\
& = \bar{g}(q) - \frac{i\pi}{2}\beta(\bar{g}(q))-\frac{\pi^2}{8}\beta'(\bar{g}(q))\beta(\bar{g}(q))+\ldots 
\end{align}
Note that since the running coupling $\bar{g}(q)$ is at most of order $\lambda$, 
we have $\beta(\bar{g}(q))\sim\lambda^2$, while $\beta'(\bar{g}(q))\sim\lambda$ and $\beta''(\bar{g}(q))\sim 1$.  It then follows that
\begin{align} \label{eq1}
\beta(\bar{g}(\rho)) &= \beta(\bar{g}(q))\, \Big[1-\frac{i\pi}{2}\beta'(\bar{g}(q))-\frac{\pi^2}{8}\big(\beta'^2(\bar{g}(q))+\beta(\bar{g}(q))\beta''(\bar{g}(q))\big)+ O(\lambda^3)\Big], 
\end{align}
from which we obtain the spectral function 
\[\label{spec_fn0}
c^{(0)}(q) = \frac{16}{\pi^2}A_0\beta^2(\bar{g}(q))\Big[1 -\frac{\pi^2}{2}\beta'^2(\bar{g}(q))-\frac{\pi^2}{4}\beta(\bar{g}(q))\beta''(\bar{g}(q))  +O(\lambda^3)\Big].
\]
This result can be checked against that obtained using the explicit solutions for the running coupling discussed in Appendix \ref{app:B}. 

In both \eqref{eq1} and \eqref{spec_fn0}, the error term $O(\lambda^3)$ is a shorthand notation collecting together terms of the form $O(\lambda^3, \lambda^2 \bar{g}(q), \lambda\bar{g}^2(q), \bar{g}^3(q))$.  Expanded out in full, we have
\begin{align}\label{spec_fn2}
c^{(0)}(q) = \frac{16}{\pi^2}A_0\beta^2(\bar{g}(q))\Big[& \big(1-\frac{\pi^2}{2}\lambda^2 +O(\lambda^3)\big)
+\big(\frac{5\pi^2}{2} b_2 \lambda+O(\lambda^2)\big)\bar{g}(q) \nn\\ & 
- \big(\frac{5\pi^2}{2} b_2^2+O(\lambda)\big)\bar{g}^2(q)+O(\bar{g}^3(q))\Big].
\end{align}

\section{Characterising the UV CFT} 
\label{AdSNorms}

At the order to which we work, the properties of the dual QFT computed above depend only on the three parameters $A_0$, $A_1$ and $A_2$ characterising the correlators \eqref{A0def}-\eqref{A2def} in the UV CFT.
Our task in this section is now to evaluate these parameters for the CFT we are interested in, namely the CFT dual to 4d Einstein gravity with a minimally coupled scalar field and a potential, as per the simplest inflationary models.

Since these parameters describe correlators in the UV CFT itself, rather than in the deformed theory, they can be evaluated simply by perturbatively solving the scalar field equation $\Box\vphi= \kappa^2 V'(\vphi)$ on a {\it fixed}  four-dimensional AdS geometry. 
An overview of these elementary 
AdS/CFT calculations is supplied in Appendix \ref{WittenDias}; 
here we simply summarise the results as we need them.

In general, the parameters $A_0$, $A_1$ and $A_2$ are related to 
the coefficients appearing in the Taylor 
expansion of the potential to quartic order,
\[
\kappa^2 V(\vphi) = -3+\frac{1}{2}m^2 \vphi^2+\frac{1}{3}g_3\vphi^3
+\frac{1}{4} g_4 \vphi^4+O(\vphi^5),
\]
where we work in units in which the AdS radius is set to one and $\kappa^2=8\pi G$.
The linear term in this expansion has been removed by shifting $\vphi$ so that the AdS critical point where $V'=0$ occurs at the origin, while for the UV dimension of $\O_0$ to be $\Delta=3-\lambda$ we require $m^2 = \lambda(\lambda-3)$.
The information in this Taylor expansion can equivalently be repackaged in terms of a superpotential $W(\vphi)$, obeying
\[
\kappa^2 V(\vphi) = \frac{1}{2}W'^2(\vphi)-\frac{3}{4}W^2(\vphi),
\]
where\footnote{We restrict our interest here to solutions for which a critical point of the potential where $V'=0$ corresponds to a critical point of the superpotential, $W'=0$.
Holographic RG flows of this type possess `fake' supersymmetry and are known to be stable both perturbatively and non-perturbatively, see \cite{Freedman:2003ax}.}
\[\label{Wexp}
W(\vphi) = -2-\frac{1}{2}\lambda\vphi^2+\frac{1}{3} w_3 \vphi^3+\frac{1}{4}w_4 \vphi^4 +O(\vphi^5).
\]
In terms of these latter coefficients, the parameters characterising the UV CFT are found in Appendix \ref{WittenDias} to be
\begin{align}\label{A0_soln}
\kappa^2 A_0 &= 1+2 b \lambda+(2b^2+4)\lambda^2+O(\lambda^3), \\[1ex]
\label{A1_soln}
\kappa^2 A_1 &=  
\frac{4w_3}{\lambda} \Big[1+3 b \lambda+(8+\frac{9b^2}{2}+\frac{\pi^2}{12})\lambda^2+O(\lambda^3)\Big], \\[1ex]
\label{A2_soln}
\kappa^2 A_2 &= \frac{20 w_3^2}{\lambda^2}\Big[1+\Big(4b-\frac{3 w_4}{10 w_3^2}\Big)\lambda+\Big(\frac{68}{5}+8 b^2-\frac{6 b w_4}{5 w_3^2}+\frac{\pi^2}{5}\Big)\lambda^2+O(\lambda^3)\Big],
\end{align} 
where
\[
b \equiv 2-\ln 2-\gamma
\]
with $\gamma$ the Euler-Mascheroni constant.    
Restored to the right-hand sides of these equations, the factor of $\kappa^{-2}$ is proportional to $N^2$, where $N$ is the rank of the gauge group of the dual QFT.
 
Using \eqref{b2b3_def}, we can then immediately read off the coefficients of $\beta(g)$:
\begin{align}\label{b2_result}
b_2 &= w_3 \Big[ 1+b\lambda+\Big(4+\frac{b^2}{2}+\frac{\pi^2}{12}\Big)\lambda^2+O(\lambda^3)\Big], \\
\label{b3_result}
b_3 &= w_4 + \frac{\lambda}{9}\Big(18 b w_4 - (48+\pi^2) w_3^2\Big) 
+ O(\lambda^2).
\end{align}
With $w_3$ and $w_4$ of order unity, $b_2$ and $b_3$ are also of order unity so the IR fixed point indeed occurs at $g_{IR}\sim \lambda$.

\section{Calibrating to the cosmology} 
\label{sec:calibrating}

In this section, as a final preparation before computing the holographic power spectrum, we refine our choice of renormalisation scheme so as to set the running coupling in the dual QFT equal to the value of the inflaton at horizon crossing in the corresponding cosmology.  

To begin, let us specify more precisely the nature of this inflationary cosmology.
As directed by the domain-wall/cosmology correspondence \cite{McFadden:2009fg,McFadden:2010na, Cvetic:1994ya,Skenderis:2006jq}, 
we consider the inflationary cosmology associated with the inverted domain-wall potential 
\[\label{Vc_def}
\kappa^2 V_c(\vphi) = - \kappa^2 V(\vphi) =  -\frac{1}{2}W'^2(\vphi)+\frac{3}{4}W^2(\vphi).
\]
The Taylor expansion of $W(\vphi)$ in \eqref{Wexp} now describes an expansion about a de Sitter fixed point.
Since by assumption our dual QFT lives on a flat metric, 
we consider spatially flat four-dimensional FRW backgrounds of the form
\[
\d s^2 = -\d t^2+a^2(t)\d \vec{x}\,^2, \qquad \vphi = \vphi(t).
\]
We will further assume the inflaton profile $\vphi(t)$ is monotonic (or at least piecewise so), as befits an RG flow.
The inflationary action
\[
S = \frac{1}{2\kappa^2}\int\d^4 x \sqrt{-g}[R-(\p\vphi)^2-2 \kappa^2 V_c(\vphi)]
\]
then yields the first-order equations of motion \cite{Salopek:1990jq}
\[\label{bgd_eom}
H \equiv \frac{\dot{a}}{a}=-\frac{1}{2}W(\vphi), \qquad \dot{\vphi} = W'(\vphi).
\]

Introducing the slow-roll parameter
\[\label{ep_def}
\ep = -\frac{\dot{H}}{H^2} = \frac{\dot{\vphi}^2}{2H^2} = \frac{2W'^2}{W^2},
\]
the log derivative of $\vphi_*(q)$, the value of the inflaton at horizon crossing 
(defined as the time at which $q=aH$), is given by
\begin{align}\label{vphi*deriv}
\frac{\d\vphi_*(q)}{\d\ln (q/\mu)} &= \frac{\sqrt{2\ep_*(q)}}{1-\ep_*(q)} 
=-\lambda\vphi_*(q)+ w_3 \vphi_*^2(q)+\big(w_4+\frac{\lambda^2}{4}-\frac{\lambda^3}{2}
\big)\vphi_*^3(q)+O(\vphi_*^4(q)),
\end{align}
where $\mu$ is again our reference momentum scale.
The QFT running coupling, on the other hand, satisfies
\[
\frac{\d\bar{g}(q)}{\d\ln (q/\mu)} = \beta(\bar{g}(q)) = -\lambda \bar{g}(q)+b_2 \bar{g}^2(q)+b_3\bar{g}^3(q)+O(\bar{g}^4(q)),
\]
with $b_2$ and $b_3$ as identified above in \eqref{b2_result}-\eqref{b3_result}. 

Comparing these equations, we see that at leading order in $\lambda$ the running coupling $\bar{g}(q)$ coincides with the value of the inflaton $\vphi_*(q)$ at horizon crossing, if we choose $g=\vphi_*(\mu)$.  (Note in our conventions the inflaton and the running coupling are both dimensionless.)
To extend this convenient identification to higher order, we modify our renormalisation scheme 
by re-defining the renormalised coupling:
\[\label{g_vphi}
g = \vphi(1+a_1 \vphi +a_2 \vphi^2+O(\vphi^3)).
\]
Under this change of scheme, the $\beta$-function becomes
\[\label{betaphi}
\beta(\vphi) = \beta(g)\Big(\frac{\d g}{\d \vphi}\Big)^{-1} = -\lambda \vphi + B_2\vphi^2+B_3\vphi^3+O(\vphi^4),
\]
where
\[\label{Bdefs}
B_2 = b_2+\lambda a_1, \qquad B_3= b_3+2\lambda a_2-2\lambda a_1^2.
\]
For $a_1$ and $a_2$ of order unity, this produces the subleading corrections to the $\beta$-function that we  require in order to match with \eqref{vphi*deriv}.
Specifically, introducing the running coupling $\bar{\vphi}(q)$ defined by
\[\label{phibardef}
\frac{\d\bar{\vphi}(q)}{\d\ln (q/\mu)} = \beta(\bar{\vphi}(q)), \qquad \bar{\vphi}(\mu) = \vphi,
\]
in order to identify 
\[\label{wB}
B_2 = w_3+O(\lambda^3), \qquad B_3 = w_4+O(\lambda^2)
\]
we must set
\begin{align}\label{a_solns1}
a_1 &= -b_2\Big(b+\big(4-\frac{b^2}{2}+\frac{\pi^2}{12}\big)\lambda\Big)+O(\lambda^2), \\[1ex]
\label{a_solns2}
a_2 &= -b \hspace{0.1mm} b_3+\Big(b^2+\frac{8}{3}+\frac{\pi^2}{18}\Big)b_2^2+O(\lambda).
\end{align}
Setting in addition $\vphi = \vphi_*(\mu)$, the running coupling $\bar{\vphi}(q)$ is now equal to the value of the inflaton at horizon crossing $\vphi_*(q)$, working\footnote{In fact, this identification can be extended to all orders in $\lambda$ by using the exact expressions for $A_0$, $A_1$ and $A_2$ (see Appendix \ref{WittenDias}) to obtain 
$b_2$ and $b_3$ in \eqref{b2_result}-\eqref{b3_result} exactly. 
One then solves for the $a_1$ and $a_2$ in \eqref{Bdefs} required to identify \eqref{betaphi} with \eqref{vphi*deriv} to all orders in $\lambda$.  All we require here however are the leading terms as given in \eqref{a_solns1}-\eqref{a_solns2}.} to order $\lambda^3$.

Under the change of renormalisation scheme \eqref{g_vphi}, the 
stress tensor 2-point function, and hence the spectral densities, are invariant.  
It will be useful, nevertheless, to re-express the spectral function $c^{(0)}(q)$   
in terms of $\beta(\bar{\vphi}(q))$ and its derivatives. 
To do so, starting from \eqref{spec_fn2} we use \eqref{g_vphi} and \eqref{betaphi} in the form
\begin{align}
\beta(\bar{g}(q)) &= \beta(\bar{\vphi}(q))\big[1+2a_1\bar{\vphi}(q)+3a_2\bar{\vphi}^2(q)+O(\bar{\vphi}^3(q))\big],\\
\qquad \bar{g}(q) &= \bar{\vphi}(q)\big[1+a_1\bar{\vphi}(q)+a_2\bar{\vphi}^2(q)+O(\bar{\vphi}^3(q))\big],
\end{align}
along with
\eqref{Bdefs}, \eqref{a_solns1} and \eqref{a_solns2}, plus the value of $A_0$ from \eqref{A0_soln}.
The spectral function then takes the form
\[\label{spec_fn}
c^{(0)}(q) = \frac{16}{\pi^2}\kappa^{-2}\beta^2\Big[1-2b\beta'+\big(4+2b^2-\frac{\pi^2}{2}\big)\beta'^2+\big(b^2-\frac{\pi^2}{12}\big)\beta''\beta+O(\lambda^3)\Big],
\]
where $\beta\equiv \beta(\bar{\vphi}(q))$ and 
primes denote derivatives with respect to the running coupling $\bar{\vphi}(q)$.

\section{The holographic power spectrum}
\label{sec:results}

Finally we are ready to determine the holographic power spectrum.
Plugging the spectral function \eqref{spec_fn} into the holographic formula \eqref{hol_form_2}, we obtain
\[\label{p1}
\Delta_S^2(q) = \frac{\kappa^2}{4\pi^2 \beta^2}\Big[1+2b\beta'+\big(-4+2b^2+\frac{\pi^2}{2}\big)\beta'^2+\big(-b^2+\frac{\pi^2}{12}\big)\beta''\beta+O(\lambda^3)\Big],
\]
where again $\beta\equiv \beta(\bar{\vphi}(q))$.
To put this result in recognisable form, we introduce the slow-roll parameters
\[
\eta = \frac{\ddot{\vphi}}{H\dot{\vphi}}=-\frac{2W''}{W}, \qquad  
\delta_2 = \frac{\dddot{\vphi}}{H^2\dot{\vphi}}=\frac{4}{W^2}(W'''W'+W''^2),
\]
in addition to $\ep$ as defined earlier in \eqref{ep_def}. 
Since we have identified $\vphi_*(q)$ with $\bar{\vphi}(q)$, it follows that
\[\label{SR_beta_eqs}
\ep_* = \frac{1}{2}\beta^2+O(\lambda^7), \qquad \eta_* = \beta'+O(\lambda^4), \qquad \delta_{2*}=\beta'^2+\beta''\beta+O(\lambda^5).
\]
The holographic power spectrum \eqref{p1} is thus
\[\label{result}
\boxed{\Delta_S^2(q) = \frac{\kappa^2 H_*^2}{8 \pi^2 \ep_*}\Big[1+2b\eta_*+(3b^2-4+\frac{5\pi^2}{12})\eta_*^2+(-b^2+\frac{\pi^2}{12})\delta_{2*
}+O(\lambda^3)\Big].}
\]
This is precisely the standard second-order slow-roll power spectrum, as we see by comparing with the conventional inflationary calculation in \cite{Gong:2001he}, noting that here
\[\label{SR_orders}
\ep_* \sim \lambda^4, \qquad \eta_* \sim \lambda, \qquad \delta_{2*}\sim \lambda^2,
\]
since $\beta\sim \lambda^2$, $\beta'\sim \lambda$ and $\beta''\sim 1$.
In writing \eqref{result} we have also used $H_*(q) = 1+O(\lambda^3)$, as follows from \eqref{bgd_eom} and \eqref{Wexp} noting that $\vphi_*(q)$ is at most of order $\lambda$.  Our choice of units in which the de Sitter radius is one originated in \eqref{Wexp}.  

The entire holographic cosmology is thus controlled by the $\beta$-function of the dual QFT.  We start by solving \eqref{phibardef} for the running coupling $\bar{\vphi}(q)$, which in our renormalisation scheme is equal to the value of the inflaton at horizon crossing $\vphi_*(q)$.   The RG scale $\mu$ plays the role of the cosmological pivot scale, with $\vphi = \bar{\vphi}(\mu)$ being the value of the inflaton when this pivot scale crosses the horizon, $\vphi_*(\mu)$.  The slow-roll parameters at horizon crossing $\ep_*$, $\eta_*$ and $\delta_{2*}$  follow from \eqref{SR_beta_eqs}, and the power spectrum is given by \eqref{result}.   
The inflaton potential $V_c(\vphi)$ can also be reconstructed from the $\beta$-function: using \eqref{Vc_def} and $-2W'(\vphi)/W(\vphi) =  \beta(\vphi) +O(\lambda^6)$, we find
\begin{align}\label{V_beta}
\kappa^2 V_c(\vphi) &=\frac{1}{2}(6-\beta^2(\vphi))\exp\Big({-}\int_0^\vphi\d\vphi\,\beta(\vphi)\Big)+O(\lambda^7) \nn\\
&=3+\frac{1}{2}\lambda(3-\lambda)-\frac{1}{3}g_3 \vphi^3-\frac{1}{4}g_4\vphi^4+O(\vphi^5) 
\end{align}
where 
\[\label{gsolns}
g_3 = 3 B_2 (1-\lambda)+O(\lambda^3), \qquad g_4 = (3-4\lambda)B_3+2B_2^2+O(\lambda^2), 
\]
setting the error terms to those in \eqref{wB}.
If desired, the background solution for $a(t)$ and $\vphi(t)$ can similarly be reconstructed from the superpotential,
\begin{align}
W(\vphi) &= -2\exp\Big(-\frac{1}{2}\int_0^\vphi \d\vphi\,\beta(\vphi)\Big)+O(\lambda^7) \nn\\
&=-2 -\frac{1}{2}\lambda\vphi^2+\frac{1}{3}(B_2+O(\lambda^3))\vphi^3+\frac{1}{4}(B_3+O(\lambda^2))\vphi^4+O(\vphi^5),
\end{align}
using the equations of motion \eqref{bgd_eom}.  (See \cite{Bzowski:2012ih} for the case where $B_3=0$.)

We emphasise that our choice of renormalisation scheme makes no impact on the final result for the cosmological power spectrum, since the spectral function $c^{(0)}(q)$ is scheme-independent.  Rather, the choice of scheme is simply a matter of convenience, determining the relationship between the running coupling and its $\beta$-function in the deformed CFT and their counterparts in the cosmology, the inflaton and the slow-roll parameters at horizon crossing.  Here we chose to identify the running coupling with the inflaton at horizon crossing, yielding the relation \eqref{SR_beta_eqs} between the slow-roll parameters at horizon crossing and the $\beta$-function of the running coupling.  An alternative choice, discussed in \cite{Larsen:2002et,Larsen:2003pf,Larsen:2004kf}, would be to identify 
$\beta(\vphi)=\sqrt{2\ep(\vphi)} = -2W'(\vphi)/W(\vphi)$, where $\vphi$ is the renormalised coupling.
In this case, however, the relation between the running coupling and the inflaton at horizon crossing would be nontrivial, $\bar{\vphi}(q) = \vphi_*(q)(1+a_2\vphi_*^2(q)+\ldots)$ with $a_2 \sim \lambda^2$.
This relation would then have to be taken 
into account in going from the spectral function to the slow-roll power spectrum.   
While these two schemes are effectively equivalent at the order to which we currently work, they will be distinct at higher order.

\begin{figure}[t]
\center
\includegraphics[height=5.2cm]{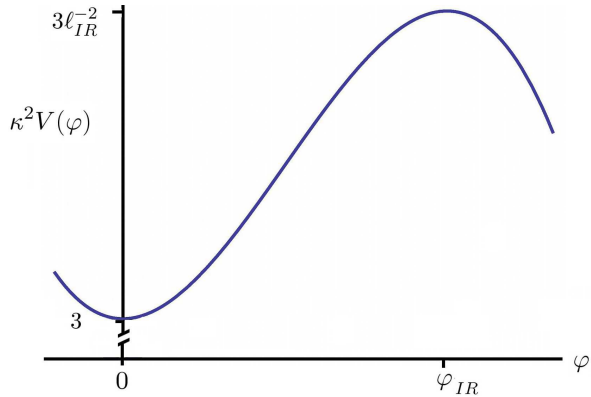} 
\caption{\label{ptl} The inflaton potential 
comprises a hilltop (corresponding to the IR fixed point at $\vphi_{IR}\sim \lambda$) with a nearby local minimum at the origin (corresponding to the UV fixed point).  The decrease in height of the potential is of order $\lambda^3$.}
\end{figure}

The form of the inflaton potential is sketched in Figure \ref{ptl}.
The UV fixed point of the dual QFT corresponds to the local minimum at the origin, while the IR fixed point  corresponds to the local maximum at $\vphi_{IR} \sim \lambda$.  Each fixed point is associated with a CFT, corresponding to a de Sitter asymptotic region in the cosmology. 
Cosmic time evolution corresponds to inverse RG flow, since the IR CFT controls the early-time behaviour and the UV CFT the late-time behaviour of the cosmology.  

The spectral index is given by\footnote{We assume here that $B_3\sim 1$ allowing us to drop a term proportional to $\beta'''\beta^2 \sim \lambda^4$. When comparing with \cite{Gong:2001he}, this corresponds to the case where $\delta_{3*} =(\ddddot{\vphi}/H^3\dot{\vphi})_* =  4\delta_{2*}\eta_*-3\eta_*^2+O(\lambda^4)$.}
\begin{align}
n_s-1 &= \frac{\d }{\d \ln q} \ln \Delta_S^2(q)=  -2\beta'+\beta''\beta -(8+b^2-\frac{13}{12}\pi^2)\beta\beta'\beta'' +O(\lambda^4) \nn\\
&=-2\eta_*+2b(\delta_{2*}-\eta_*^2)-(8+b^2-\frac{13}{12}\pi^2)(\delta_{2*}-\eta_*^2)\eta_*+O(\lambda^4).
\end{align}
At the UV and IR fixed points, however, $\beta$ vanishes and we have simply
\[
(n_s-1)|_{UV} = -2\beta'_{UV} = 2\lambda, \qquad 
(n_s-1)|_{IR} = -2\beta'_{IR} = -2\lambda-\frac{2B_3}{B_2^2}\lambda^2+O(\lambda^3).
\]
Short wavelength modes probing the UV fixed point thus see a slightly blue-tilted spectrum while long wavelength modes controlled by the IR fixed point experience a slight red tilt. To obtain $n_s \approx 0.96$ in the IR requires $\lambda \approx 0.02$, consistent with 
our assumption that $\lambda \ll 1$.  
Similarly, the observed small amplitude of the primordial power spectrum $\Delta_S^2\sim 10^{-9}$ implies $\kappa^2 \lambda^{-4} \sim  10^{-7}$ in our units where the de Sitter radius is one.  Since $\kappa^{2}\sim N^{-2}$, where $N$ is the rank of the gauge group of the dual QFT, this then implies $N\sim 10^6$  consistent with the large-$N$ limit. 

As mentioned in the introduction, our present scenario does not yet include a mechanism for inflation to end.  A natural way to describe this holographically would be to relax our assumption that the RG flow is driven by a single scalar operator, corresponding to a single scalar field in the cosmology.  The inflaton would then 
be interpreted as 
the most nearly marginal of the irrelevant operators about the IR fixed point, since this operator controls the behaviour nearest the IR fixed point, corresponding to the earliest times in the cosmology.  The other operators would  become important around the time of reheating, altering the nature of the RG flow in the UV.

\section{Discussion}
\label{sec:discussion}

In this paper we calculated the power spectrum of inflationary cosmologies that have a dual description in terms of a deformed CFT.  Our calculations were based primarily in the dual QFT: using second-order conformal perturbation theory we determined the spectral density associated with the 2-point function for the trace of the stress tensor.  This quantity is inversely proportional to the cosmological power spectrum, as follows 
from the holographic framework of \cite{McFadden:2009fg}.
To calibrate the renormalisation scheme of our dual QFT to the bulk cosmology, we additionally supplied three input parameters $A_0$, $A_1$ and $A_2$ obtained via standard AdS/CFT calculations on a fixed (i.e., non-dynamical) AdS background.  These parameters characterise soft limits of 2-, 3- and 4-point correlators in the UV CFT, and serve to connect the $\beta$-function in the dual QFT to the asymptotic form of the inflaton potential.  Knowing this relation allows us to define the off-critical renormalised scalar operator in the dual QFT so that its running coupling coincides with the bulk inflaton at horizon crossing.  With this choice of scheme, it is then simple to express the holographic power spectrum obtained from conformal perturbation theory in terms of the bulk inflationary slow-roll parameters.  The result we obtain exactly reproduces the standard inflationary power spectrum to second order in slow roll, working 
in the appropriate limit where $\ep\ll \eta$.
In combination with the leading-order results of \cite{Bzowski:2012ih} for non-Gaussianities, our calculations 
provide a highly non-trivial successful test of holographic cosmology.

Our analysis can readily be extended 
in a number of interesting directions.
The most obvious is the calculation of non-Gaussianities beyond leading order in conformal perturbation theory: this is important because various pieces of correlators, for example the equilateral piece of the scalar bispectrum, do not appear until higher order \cite{Bzowski:2012ih}.  Moreover other correlators, such as that of two gravitons and a scalar, are zero at leading order in conformal perturbation theory.
On a similar note, the accuracy of our present results could also easily be improved to $O(\bar{\vphi}^3(q))$, rather than $O(\lambda^3)$ as given here.  This would require using the exact expressions for $A_0$, $A_1$ and $A_2$, rather than the series expansions in $\lambda$ we have used here, and the exact solution for $\bar{g}(\rho)$ in \eqref{grhoeq}.  Since the running coupling $\bar{\vphi}(q)\ll \lambda$ in the vicinity of the UV fixed point, this would give greater accuracy at high momenta.  We have not pursued this refinement here however, partly in the interests of simplicity, and partly because the red tilt of the observed power spectrum implies the momenta relevant to the CMB lie near the IR fixed point where the running coupling $\bar{\vphi}(q)\sim \lambda$.  For these small momenta there would thus 
be little 
improvement in accuracy.   More generally, the extension of our calculation to higher orders in conformal perturbation theory also appears quite feasible, with the only significant penalty incurred being the 
calculation of 
extra Witten diagrams beyond those in Appendix \ref{WittenDias}.  One motivation for these efforts would be to recover holographically the slow-roll corrections proportional to powers of $\ep$.

A more ambitious and intriguing question is whether our conformal perturbation theory methods can be extended to the case where the slow-roll parameters $\ep$ and $\eta$ are of similar magnitude.  
In essence, our present scenario has $\ep\ll \eta$ because the cosmology is undergoing a slow-roll between two nearby extrema of the inflationary potential, and at each extremum $\ep$ vanishes.  In the dual QFT, this corresponds to the close proximity of the UV and IR fixed points in the space of couplings.  To obtain $\ep\sim \eta$ for part of the RG trajectory requires moving these two fixed points further apart (or considering the flow to the other zero of our cubic $\beta$-function, located at unit distance), but it appears difficult to do this without losing control over the conformal perturbation theory.  The problem seems reminiscent 
of that of analysing critical behaviour in three dimensions via the $\ep$-expansion, 
where the separation of the Gaussian and Wilson-Fisher fixed points is of order $\ep$ in $d=4-\ep$ dimensions.
From this perspective it may be worthwhile 
to explore the various resummation techniques developed in this latter context.

Besides the case studied here of a CFT deformed by a slightly relevant scalar operator, it is also interesting to consider 
the case of a CFT deformed by a marginally relevant scalar operator with a 
$\beta$-function of the form
\[\label{YMbeta}
\beta(g) = -b_2 g^2+ b_3 g^3+O(g^4).
\]
Indeed, from the perspective of minimising the amount of fine tuning, the case $\lambda=0$ may be preferable to that studied here with a small parameter $\lambda \ll 1$.  
Our present methods should nevertheless be straightforwardly applicable, leading to a cosmological power spectrum with logarithmic rather than power-law deviations from scale invariance.  
In fact, a power spectrum of precisely this form already appeared in \cite{McFadden:2010na, Easther:2011wh}, based on holographic models in which the dual QFT was taken to be $SU(N)$ Yang-Mills theory plus adjoint matter. In the weak coupling regime studied, the $\beta$-function for these models takes the form \eqref{YMbeta}.  Holographic cosmologies in which the dual QFT is asymptotically free have also recently been discussed in \cite{Kiritsis:2013gia}.

Other interesting open directions include the 
holographic modelling of reheating and the analysis of
multiscalar cosmologies, in which the generation of entropy perturbations should map to the mixing of operators under RG flow in the dual QFT. 
Such cosmologies also  
offer the possibility of crossover behaviour in which the RG flow approaches close to the UV/IR fixed point but then turns away along a new relevant direction, so that the cosmology is not necessarily asymptotically de Sitter all the way into the future/past.
Finally, we might wonder whether our explicit dual holographic description of a hilltop inflationary potential can offer any insight into the nature of the eternal inflationary regime.

\bigskip

{\small
{\it Acknowledgments:}  
We thank Kostas Skenderis and Adam Bzowski 
for discussions and comments on the manuscript. 
Research at the Perimeter Institute is supported by the Government of Canada
through Industry Canada and by the Province of Ontario through the Ministry of
Research and Innovation.
}

\appendix

\section{Solution for the running coupling}
\label{app:B}

In this appendix, we solve \eqref{phibardef} for the running coupling $\bar{\vphi}(q)$ in the deformed CFT, taking the opportunity to verify the consistency of our results with those of 
\cite{Bzowski:2012ih}.
We first consider the case of a purely quadratic $\beta$-function of the form \eqref{betaphi} with $B_3$ set to zero, as studied in \cite{Bzowski:2012ih} at leading order in conformal perturbation theory.  
Solving for the running coupling, we obtain
\[
\bar{\vphi}(q) = \vphi_1 [1+(q/q_0)^\lambda]^{-1}, 
\]
where $\vphi_1=\lambda/B_2$ is the location of the IR fixed point and   
\[
q_0^{-\lambda} = \Big(\frac{\vphi_1}{\vphi}-1\Big)\mu^{-\lambda}.
\]
We then obtain a power spectrum of the form \eqref{result}, where from \eqref{SR_beta_eqs}
\begin{align}
\ep_* &= \frac{1}{2}\lambda^2\vphi_1^2 (q/q_0)^{2\lambda}[1+(q/q_0)^\lambda]^{-4}\big[1+O(\lambda^3)\big], \nn\\
\eta_*& = \lambda \big[-1+2[1+(q/q_0)^\lambda]^{-1}+O(\lambda^3)\big], \nn\\
\delta_{2*} &= \lambda^2\big[1-6[1+(q/q_0)^\lambda]^{-1}+6[1+(q/q_0)^\lambda]^{-2}+O(\lambda^3)\big].
\end{align}
The leading term in this power spectrum indeed agrees with that found in \cite{Bzowski:2012ih}.   (See, e.g., Sections 4.2 and 4.3 of \cite{Bzowski:2012ih}. Our $q_0$ here matches that in \cite{Bzowski:2012ih}, since  $\bar{\vphi}(q)$ is equal to $\vphi_*(q)$, and our $\vphi_1$ agree since  $B_2$ here is $2\pi C$ there.)
 
We can similarly verify that the result quoted for the scalar 2-point function in \cite{Bzowski:2012ih}
(the final equation in Section 2.2), 
agrees with the leading term in $\lla \O_0(q)\O_0(-q)\rra$ here, to which it corresponds.  From \eqref{renOdef}, \eqref{beta1} and \eqref{scalar2ptresult}, we have
\[
\lla\O_0(q)\O_0(-q)\rra = \frac{A_0 q^3}{\lambda^2\phi_0^2}\beta^2(\bar{g}(q)).
\]
At leading order, $\beta^2(\bar{g}(q))=\beta^2(\bar{\vphi}(q)) = 2\ep_*(q)$ while $A_0 = \kappa^{-2}$, and so we indeed recover the result of \cite{Bzowski:2012ih}, noting that $\phi_0 = \vphi_1 q_0^{-\lambda}$ and that $\alpha=12/\pi^2\kappa^2$ there.
 
Let us now turn to the problem of solving for the running coupling $\bar{\vphi}(q)$ when $B_3$ is nonzero.  In this case, integrating \eqref{phibardef} with the $\beta$-function \eqref{betaphi}, we obtain the implicit solution 
\[\label{implicit}
\Big(\frac{q}{\mu}\Big)^\lambda = \phi \Big(\frac{1}{\bar{\vphi}(q)}-\frac{1}{\vphi_+}\Big)^{\frac{\vphi_-}{\vphi_- - \vphi_+}}
\Big(\frac{1}{\bar{\vphi}(q)}-\frac{1}{\vphi_-}\Big)^{\frac{\vphi_+}{\vphi_+ - \vphi_-}}
\]
where
\[\label{phipm}
\vphi_\pm = \frac{B_2}{2B_3}\Big[-1\pm\sqrt{1+\frac{4\lambda B_3}{B_2^2}}\Big]
\]
are the two nonzero roots of the $\beta$-function, and $\phi$ is chosen so that $\bar{\vphi}(\mu)=\vphi$, namely
\[\label{phisol}
\phi = \Big(\frac{1}{\vphi}-\frac{1}{\vphi_+}\Big)^{\frac{\vphi_-}{\vphi_+ - \vphi_-}}
\Big(\frac{1}{\vphi}-\frac{1}{\vphi_-}\Big)^{\frac{\vphi_+}{\vphi_- - \vphi_+}}.
\]

While \eqref{implicit} cannot be inverted to give an exact solution for $\bar{\vphi}(q)$, we can obtain 
an iterative solution to the required order,  
\begin{align}
\bar{\vphi}(q) &= \frac{\lambda}{B_2} \frac{\theta}{1+\theta} \Big[ 1+ \frac{\lambda B_3}{B_2^2}\frac{\theta}{1+\theta}\Big(\ln(1+\theta)-\theta\Big) \nn\\ \qquad
& + \frac{\lambda^2 B_3^2}{B_2^4}\frac{1}{(1+\theta)^2}\Big(3\theta+2\theta^2-(4\theta+3)\ln(1+\theta)
+\frac{1}{2}(1-\theta)\ln^2 (1+\theta)\Big) + O(\lambda^3)\Big]
\end{align}
where
\[
\theta = \frac{\phi B_2}{\lambda}\Big(\frac{q}{\mu}\Big)^{-\lambda}.
\]
Note that the prefactor here is actually of order unity, since $\phi\sim\lambda$ from \eqref{phisol} and \eqref{phipm}.
When Taylor expanded about the UV as $\bar{\vphi}(q) = \lambda \sum_{n=1}^\infty c_n \theta^n$, this iterative solution captures the correct behaviour of all the order one coefficients $c_n$ up to an error $O(\lambda^3)$.
(This can be checked explicitly against the solution of \eqref{phibardef} about the UV fixed point.)
If desired, 
the slow-roll parameters at horizon crossing can now be evaluated 
as functions of momentum using \eqref{SR_beta_eqs}.

\section{Correlators in the UV CFT} 
\label{WittenDias}

In this appendix, we use the AdS/CFT correspondence \cite{Maldacena:1997re,Gubser:1998bc,Witten:1998qj,Aharony:1999ti} 
to evaluate the parameters $A_0$, $A_1$ and $A_2$ appearing in the UV CFT correlators \eqref{A0def}-\eqref{A2def}. 
First, we perturbatively solve the equation of motion for an interacting scalar field on a fixed AdS background. 
Correlators in the UV CFT can then be read off using the holographic prescription of \cite{deHaro:2000xn}. 
Since computations of this nature 
are by now standard, our treatment here will be brief; 
for an introductory review with 
links to the original literature 
we refer the reader to, e.g., \cite{Skenderis:2002wp}.
 
Starting from the Taylor expansion of the superpotential about the UV fixed point given in \eqref{Wexp},
\[
W(\vphi) = -2-\frac{1}{2}\lambda\vphi^2+\frac{1}{3}w_3 \vphi^3 + \frac{1}{4}w_4\vphi^4+O(\vphi^5),
\]
the corresponding potential reads  
\[
\kappa^2 V(\vphi) = \frac{1}{2}W'\,^2(\vphi)-\frac{3}{4}W^2(\vphi) = 
-3+\frac{1}{2}m^2 \vphi^2+\frac{1}{3}g_3\vphi^3
+\frac{1}{4} g_4 \vphi^4+O(\vphi^5)
\]
where
\[\label{g_defs}
m^2 = \lambda(\lambda-3), \qquad g_3 = 3 w_3(1-\lambda), \qquad 
g_4 = 2w_3^2 -\frac{3}{4}\lambda^2+w_4(3-4\lambda).
\]

To compute the required correlation functions it suffices to treat $\vphi$ 
as a small perturbation, 
i.e., we expand
\[
\vphi = \nu \vphi_{[0]}+\nu^2\vphi_{[1]}+\nu^3 \vphi_{[2]}+\O(\nu^4), \qquad \nu \ll 1.
\]
From the scalar field equation
\[
0= (-\Box+m^2)\vphi+g_3\vphi^2+g_4\vphi^3+O(\vphi^4),
\]
we then have
\begin{align}
0 &= (-\Box+m^2)\vphi_{[0]}, \\
0 &= (-\Box+m^2)\vphi_{[1]}+g_3\vphi_{[0]}^2,\\
0 &= (-\Box+m^2)\vphi_{[2]}+2g_3\vphi_{[0]}\vphi_{[1]}+g_4\vphi_{[0]}^3.
\end{align}
Working on a four-dimensional AdS background with metric $\d s^2 = z^{-2}[\d z^2+\d\vec{x}\,^2]$, we find the bulk-boundary propagator
\[
\mathcal{K}_q(z) = \frac{q^{3/2-\lambda}z^{3/2}}{2^{1/2-\lambda}\Gamma(3/2-\lambda)}K_{3/2-\lambda}(qz),
\]
normalised so that $\mathcal{K}_q(z)\tto 1$ as $z\tto 0$,
and the bulk-bulk propagator
\[ 
\mathcal{G}_q(z,z') = \begin{cases}
(z z')^{3/2} K_{3/2-\lambda}(qz)I_{3/2-\lambda}(qz') & z>z' \\
(z z')^{3/2} I_{3/2-\lambda}(qz)K_{3/2-\lambda}(qz') & z'>z. 
\end{cases}
\] 
where $(-\Box+m^2)\mathcal{G}_q(z,z')= z^{-4}\delta(z-z')$.

We then obtain the solutions
\begin{align}\label{phi0sol}
\vphi_{[0]}(z,\q_1) &= \mathcal{K}_q(z)\phi_{(0)}(\q_1), \\
\label{phi1sol}
\vphi_{[1]}(z,\q_1) &= -g_3\int[[\d p_1\d p_2]]\phi_{(0)}(-\vec{p}_1)\phi_{(0)}(-\vec{p}_2)
\int\frac{\d z'}{z'^4}\mathcal{G}_{q_1}(z,z')\mathcal{K}_{p_1}(z')\mathcal{K}_{p_2}(z'),\\
\label{phi2sol}
\vphi_{[2]}(z,\q_1) &= \int[[\d p_1\d p_2\d p_3]]\phi_{(0)}(-\vec{p}_1)\phi_{(0)}(-\vec{p}_2)\phi_{(0)}(-\vec{p}_3) \nn\\
&\qquad  \Big[
-g_4\int \frac{\d z'}{z'^4} \mathcal{G}_{q_1}(z,z')\mathcal{K}_{p_1}(z')\mathcal{K}_{p_2}(z')\mathcal{K}_{p_3}(z') \nn\\&\qquad
+2g_3^2 \int\frac{\d z'}{z'^4}\int\frac{\d z''}{z''^4} \mathcal{G}_{q_1}(z,z')\mathcal{K}_{p_1}(z')\mathcal{G}_{\sqrt{q_1^2+p_1^2}}(z',z'')\mathcal{K}_{p_2}(z'')\mathcal{K}_{p_3}(z'')\Big] 
\end{align}
where the momentum space integration measure
\[
[[\d p_1 \ldots \d p_n]] = (2\pi)^3\delta(\q_1+\sum_{i=1}^n \vec{p}_i).
\]

Asymptotically, as $z\tto 0$ the scalar field behaves as
\[
\vphi = z^\lambda\Big[\phi_{(0)} + z^2\phi_{(2)}+\ldots + z^{3-2\lambda}\phi_{(3-2\lambda)}+\ldots\Big].
\]
Up to local contact terms, the 1-point function in the presence of sources is then
\[
\<\O_0\>_s = -(3-2\lambda)\kappa^{-2}\phi_{(3-2\lambda)}.
\]
Since in the limit $z\tto 0$, with $z'>z$, 
\begin{align}
\mathcal{G}_{q_1}(z,z') &= z'^{3/2}K_{3/2-\lambda}(q_1 z') \Big[\frac{q_1^{3/2-\lambda}}{2^{3/2-\lambda}\Gamma(5/2-\lambda)} z^{3-\lambda}+O(z^{5-\lambda})\Big] \nn\\ 
&= \frac{1}{(3-2\lambda)}\mathcal{K}_{q_1}(z') z^{3-\lambda} + O(z^{5-\lambda}),
\end{align}
the contributions of $\vphi_{[1]}$ and $\vphi_{[2]}$ to 
$\<\O_0(\q_1)\>_s$ are given by the solutions \eqref{phi1sol} and \eqref{phi2sol} with $G_{q_1}(z,z')$ replaced by $-\mathcal{K}_{q_1}(z')$.  

Correlation functions are obtained by differentiating the 1-point function with respect to the source, e.g.,
\[
\< \O_0(\q_1)\O_0(\q_2)\O_0(\q_3)\O_0(\q_4)\>_0 = -\frac{\delta \<\O_0(\q_1)\>_s}{\delta\phi_{(0)}(-\q_2)\delta\phi_{(0)}(-\q_3)\delta\phi_{(0)}(-\q_4)}\Big|_{\phi_{(0)}=0}.
\]
At linear order, we obtain a 2-point function of the form \eqref{A0def} with 
\[
A_0 = \frac{(3-2\lambda)\Gamma(-3/2+\lambda)}{2^{3-2\lambda}\Gamma(3/2-\lambda)}\kappa^{-2},
\]
whose series expansion in $\lambda$ is given in \eqref{A0_soln}.
In position space, this is equivalent to the normalisation
\[
\<\O_0(x_1)\O_0(x_2)\>_0 = \frac{(3-2\lambda)\Gamma(3-\lambda)}{\pi^{3/2}\Gamma(3/2-\lambda)} \frac{\kappa^{-2}}{|x_{12}|^{2\Delta}}.
\]
Next, we find the 3-point function 
\[
\lla\O_0(q)\O_0(-q)\O_0(0)\rra_0 = A_1 q^{3-3\lambda} = 2 g_3\kappa^{-2} \int \frac{\d z'}{z'^4} \mathcal{K}^2_{q}(z')\mathcal{K}_0(z').
\]
Using $\mathcal{K}_0(z')=z'^\lambda$ and the integral
\[\label{2Kint}
\int_0^\infty \d x \, x^{2\alpha-1}K_\nu^2(x) = \frac{2^{2\alpha-3}}{\Gamma(2\alpha)}\Gamma^2(\alpha)\Gamma(\alpha+\nu)\Gamma(\alpha-\nu),
\]
we obtain
\[
A_1 =g_3\kappa^{-2}\, \frac{\Gamma^2(\lambda/2)\Gamma(3(\lambda-1)/2)\Gamma((3-\lambda)/2)}{2^{3-3\lambda}\Gamma^2(3/2-\lambda)\Gamma(\lambda)}.
\]
Replacing $g_3=3w_3(1-\lambda)$ and expanding in $\lambda$ then yields \eqref{A1_soln}.
In position space, writing $x_i-x_j=x_{ij}$, this is equivalent to
\[
\<\O_0(x_1)\O_0(x_2)\O_0(x_3)\>_0 = g_3 \kappa^{-2}  \frac{\Gamma(3-3\lambda/2)\Gamma^3((3-\lambda)/2)}{\pi^3 \Gamma^3(3/2-\lambda)}\,\frac{1}{|x_{12}|^\Delta |x_{23}|^\Delta |x_{31}|^\Delta}.
\]

\begin{figure}[t]
\center
\includegraphics[height=4.5cm]{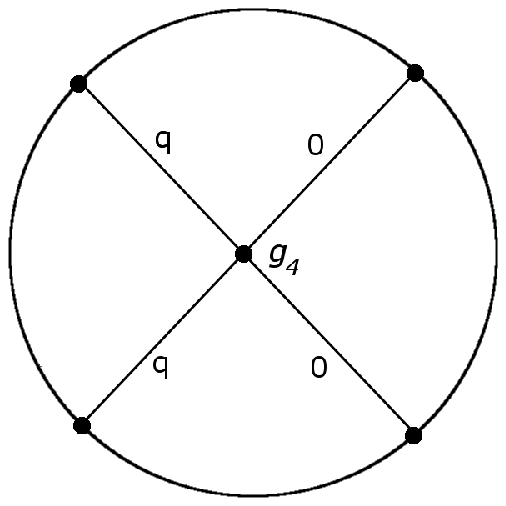} 
\includegraphics[height=4.6cm]{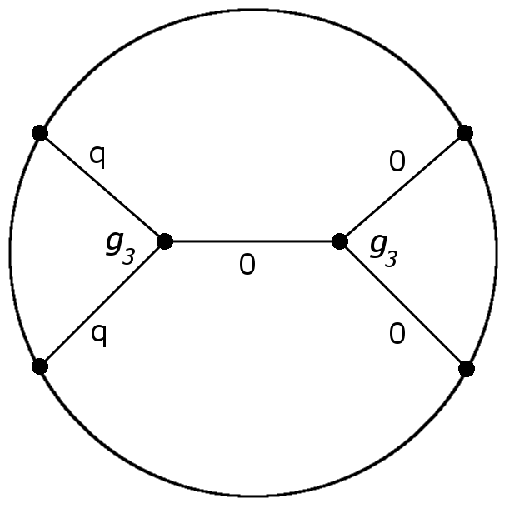} 
\includegraphics[height=4.6cm]{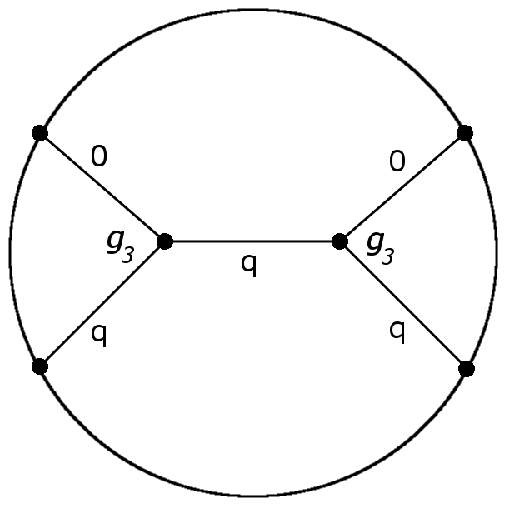} 
\caption{\label{Witten_fig} Witten diagrams contributing to $A_2$.}
\end{figure}

Finally, for the 4-point function, we find
\begin{align}
&\lla \O_0(\q_1)\O_0(\q_2)\O_0(\q_3)\O_0(\q_4)\rra_0 
= -6 g_4 \kappa^{-2}\int \frac{\d z'}{z'^4} \prod_{i=1}^4 \mathcal{K}_{q_i}(z') \nn\\&\qquad\qquad
+2 g_3^2 \kappa^{-2}\int\frac{\d z'}{z'^4}\int \frac{\d z''}{z''^4}\Big[2\mathcal{K}_{q_1}(z')\mathcal{K}_{q_2}(z')G_{\sqrt{q_1^2+q_2^2}}(z',z'')\mathcal{K}_{q_3}(z'')\mathcal{K}_{q_4}(z'') \nn\\&\qquad\qquad\qquad\qquad\qquad\qquad\qquad\qquad\qquad\qquad + (\mathrm{cyclic\, perms\, of\,}q_2,\,q_3,\,q_4) \Big],
\end{align}
and hence
\begin{align}
&\lla \O_0(q)\O_0(-q)\O_0(0)\O_0(0)\rra_0 
= -6 g_4\kappa^{-2} \int \frac{\d z'}{z'^4} \mathcal{K}_{q}^2(z')\mathcal{K}_0^2(z') \nn\\&\quad 
+4 g_3^2 {\kappa^{-2}}\int\frac{\d z'}{z'^4}\int \frac{\d z''}{z''^4}\Big[\mathcal{K}_{q}^2(z')\mathcal{G}_0(z',z'')\mathcal{K}_0^2(z'') 
+2\mathcal{K}_q(z')\mathcal{K}_0(z')\mathcal{G}_q(z',z'')\mathcal{K}_q(z'')\mathcal{K}_0(z'')\Big].
\end{align}
The three contributing diagrams are illustrated in Figure \ref{Witten_fig}.
Noting the bulk-bulk propagator at zero momentum is
\[
\mathcal{G}_0(z',z'') = \frac{1}{3-2\lambda}\big[ z'^\lambda z''^{3-\lambda}\Theta(z'-z'')
+ z''^\lambda z'^{3-\lambda}\Theta(z''-z')\big],
\]
after some straightforward algebra we recover \eqref{A2def} 
with  
\[\label{A4}
A_2 = \frac{\kappa^{-2}}{2^{1-2\lambda}\Gamma^2(3/2-\lambda)}\Big[\Big(\frac{4g_3^2}{3\lambda(1-\lambda)}-6g_4\Big)\I_1
+16 g_3^2 \I_2\Big]. 
\]

Here, the integrals
\begin{align}\label{I1_result}
\I_1 &= \int_0^\infty \d x\, x^{2\lambda-1}K_{3/2-\lambda}^2(x) = \frac{\sqrt{\pi}\Gamma^2(\lambda)\Gamma(2\lambda-3/2)}{4^{2-\lambda}\Gamma(2\lambda)}, \\
\I_2 &= \int_0^\infty \d x\, x^{\lambda-1} K_{3/2-\lambda}^2(x)\int_0^x\d x'\, x'^{\lambda-1}K_{3/2-\lambda}(x')I_{3/2-\lambda}(x').
\end{align}
To evaluate the second integral $\I_2$, we use
\[
K_\nu(x') = \frac{\pi}{2\sin \pi\nu}(I_{-\nu}(x')-I_\nu(x'))
\]
and the series expansion
\[
I_{\mu}(x')I_{\nu}(x') = \Big(\frac{x'}{2}\Big)^{\mu+\nu}\sum_{n=0}^\infty \frac{\Gamma(\mu+\nu+2n+1)}{n!\, \Gamma(\mu+\nu+n+1)\Gamma(\mu+n+1)\Gamma(\nu+n+1)}\Big(\frac{x'^2}{4}\Big)^n
\]
(see, e.g., DLMF 10.31.3) to perform the integral over $x'$.
The integral over $x$ may then be performed using \eqref{2Kint}
leading to the result
\begin{align}\label{big_result}
\I_2 = \frac{\pi}{8\cos(\pi\lambda)}\sum_{n=0}^\infty\frac{\Gamma(n+\lambda)\Gamma(n+3/2)}{(2n+\lambda)\, n!\,\Gamma(n+5/2-\lambda)}\Big[&\frac{(2n+\lambda)}{(2n+3-\lambda)}\frac{\Gamma(n+3-\lambda)\Gamma(n+2-\lambda)}{\Gamma(n+4-2\lambda)\Gamma(n+2)} \nn\\ & -\frac{\Gamma(n+1/2)\Gamma(n-3/2+2\lambda)}{\Gamma(n+1/2+\lambda)\Gamma(n-1/2+\lambda)}\Big].
\end{align}
This expression can in fact be resummed in terms of  $_pF_q$ generalised hypergeometric functions.
Here, however, we only require the first three terms of the series expansion in $\lambda$. 
The $n=0$ term in the sum evaluates to
\[
\frac{\pi}{18\lambda^2}\Big[1+(\frac{13}{6}+2b)\lambda+(401+72 b^2+156 b+30 \pi^2)\frac{\lambda^2}{36}+O(\lambda^3)\Big].
\]
The contribution from the remaining terms in the sum then begins at order unity. This order unity contribution, which is all we need, may be evaluated simply by setting $\lambda=0$ and summing over $n=1\ldots \infty$, yielding
\[
\frac{\pi}{8}\sum_{n=1}^\infty \frac{-6(4n+3)}{n^2(2n+3)^2(2n-3)(n+3)} = \frac{\pi}{216}(29-2\pi^2).
\]
To the required order, we thus have
\[\label{I2_result}
\I_2 = \frac{\pi}{18\lambda^2}\Big[1+(\frac{13}{6}+2b)\lambda+(122+18 b^2+39 b+6 \pi^2)\frac{\lambda^2}{9}+O(\lambda^3)\Big].
\]
Inserting the results \eqref{I1_result} and \eqref{I2_result} for $\I_1$ and $\I_2$ into our expression \eqref{A4} for $A_2$ and making use of \eqref{g_defs}, we recover the result quoted in \eqref{A2_soln},
\begin{align}
\kappa^2 A_2 &= \frac{20 g_3^2}{9\lambda^2}+\frac{1}{\lambda}\Big(\frac{4}{9}(11+20 b)g_3^2-2g_4\Big) \nn\\ &\qquad +\frac{4}{27}\Big((259+12b(11+10b)+3\pi^2)g_3^2-18(1+3b)g_4\Big)+O(\lambda)\nn\\[1ex]
&= \frac{20w_3^2}{\lambda^2}\Big[1+\Big(4b-\frac{3 w_4}{10 w_3^2}\Big)\lambda+\Big(\frac{68}{5}+8 b^2-\frac{6 b w_4}{5 w_3^2}+\frac{\pi^2}{5}\Big)\lambda^2+O(\lambda^3)\Big].
\end{align}

\bibliography{Def1}
\bibliographystyle{jhep}

\end{document}